\documentclass[sigconf]{acmart}
\AtBeginDocument{%
  }

\setcopyright{acmlicensed}
\copyrightyear{2018}
\acmYear{2018}
\acmDOI{XXXXXXX.XXXXXXX}
\acmConference[Tools for Thought Workshop]{Tools for Thought Workshop at CHI 2025}{CHI 2025}{Yokohama, Japan}

\acmISBN{978-1-4503-XXXX-X/2018/06}

\usepackage{array}
\usepackage{listings}
\usepackage{xcolor}

\newcommand\edit[1]{\textcolor{black}{#1}}

\definecolor{codegreen}{rgb}{0,0.6,0}
\definecolor{codegray}{rgb}{0.5,0.5,0.5}
\definecolor{codepurple}{rgb}{0.58,0,0.82}
\definecolor{backcolour}{rgb}{0.95,0.95,0.92}
 
\lstdefinestyle{mystyle}{
    backgroundcolor=\color{backcolour},   
    commentstyle=\color{codegreen},
    keywordstyle=\color{magenta},
    numberstyle=\tiny\color{codegray},
    stringstyle=\color{codepurple},
    basicstyle=\footnotesize,
    breakatwhitespace=false,         
    breaklines=true,                 
    captionpos=b,                    
    keepspaces=true,                 
    numbers=left,                    
    numbersep=5pt,                  
    showspaces=false,                
    showstringspaces=false,
    showtabs=false,                  
    tabsize=2
}

\begin{document}

\title{Who's the Leader? Analyzing Novice Workflows in LLM-Assisted Debugging of Machine Learning Code}

\author{Jessica Y. Bo}
\affiliation{%
  \institution{University of Toronto}
  \city{Toronto}
  \country{Canada}
  }
\email{jbo@cs.toronto.edu}

\author{Majeed Kazemitabaar}
\affiliation{%
  \institution{University of Toronto}
  \city{Toronto}
  \country{Canada}
  }
\email{majeed@dgp.toronto.edu}

\author{Emma Zhuang}
\affiliation{%
  \institution{University of Toronto}
  \city{Toronto}
  \country{Canada}
  }
\email{jianyun.zhuang@mail.utoronto.ca}

\author{Ashton Anderson}
\affiliation{%
  \institution{University of Toronto}
  \city{Toronto}
  \country{Canada}
  }
\email{ashton@cs.toronto.edu}

\renewcommand{\shortauthors}{Bo et al.}

\begin{abstract}
  While LLMs are often touted as tools for democratizing specialized knowledge to beginners, their actual effectiveness for improving task performance and learning is still an open question.  It is known that novices engage with LLMs differently from experts, with prior studies reporting meta-cognitive pitfalls that affect novices' ability to verify outputs and prompt effectively. 
  We focus on a task domain, machine learning (ML), which embodies both high complexity and low verifiability to understand the impact of LLM assistance on novices. Provided a buggy ML script and open access to ChatGPT, we conduct a formative study with eight novice ML engineers to understand their reliance on, interactions with, and perceptions of the LLM. We find that user actions can be roughly categorized into \textit{leading the LLM} and \textit{led-by the LLM}, \edit{and further investigate how they affect reliance outcomes like over- and under-reliance. These results have implications on novices' cognitive engagement in LLM-assisted tasks and potential negative effects on downstream learning. } 
  Lastly, we pose potential augmentations to the novice-LLM interaction paradigm to promote cognitive engagement.
\end{abstract}



\maketitle

\section{Introduction}
\label{sec:intro}

Large Language Models (LLMs) have shown remarkable ability to achieve exemplary performance in a variety of tasks, including coding, writing, and standardized tests \cite{zheng2023survey, ghazal2013bigbench}. 
However, their effectiveness in adapting to diverse end-users, ranging in their abilities in the task, still has room for improvement \cite{chen2024learning, diaz2024helping, nguyen2024beginning, lucchetti2024substance}.
In the idealized human-LLM collaboration scenario, assistance provided by LLMs should not only help with automating manual work, but also democratizes access to task-specific knowledge for novices and laypeople \cite{wang2024bridging}.
Despite these aspirational goals, this partnership breaks down when the users --- who are supposed to be in charge --- are not actually experts in the tasks. For example, prior research in the LLM-assisted coding domain has documented the meta-cognitive struggles that novices face in prompting and verifying LLMs, leading to ‘rabbitholes’ of over-reliance \cite{kazemitabaar2023novices, tie2024llms, prather2023s, lucchetti2024substance, prather2024widening}. 

We hypothesize that novice-LLM communication barriers increase in tasks involving highly complex relationships. In domains where the solution is not as easily verifiable as an introductory programming assignment, people must remain cognitively engaged in the task even whilst working alongside powerful LLM partners. To give a concrete example, machine learning (ML) is such a domain where the relationship between input hyperparameters and output metrics is not easily predictable \cite{amershi2019software, krishnan2017palm}. There is not one direct path towards the optimal solution, but a complex and iterative process that involves constant verification \cite{nahar2023meta, ashmore2021assuring, arteaga2024support}. Expert implementers can apply their learned domain-specific ML diagnostic skills, but novices lack this \cite{amershi2015modeltracker, schoop2021umlaut, yang2018grounding}. This opens up a space for LLMs to lend support to novices \cite{arteaga2024support, cao2023study}.

In this paper, we investigate what problems ML novices, with poor mental models and little implementation experience, face in an LLM-assisted workflow. We conduct a formative study with eight participants to understand how they rely on ChatGPT in a challenging ML debugging task. We identify two types of reliance actions (\textit{leading} and being \textit{led-by }the LLM) that result in different reliance outcomes, including different categories of usage errors. Our results suggest that even among novices, prior ML skill levels may correlate with reliance actions and task performance. We pose discussion around the broader consequences of using LLMs to scaffold novices’ learning, and how the novice-LLM interaction framework can be augmented to improve novices’ mental models and prevent over-reliance on suboptimal LLM responses.

\section{Formative Study}
\label{sec:method}

We conducted an in-person formative study at a research university in Canada. We recruited eight participants from the undergraduate and graduate student body through Slack channels and word-of-mouth. Participants were asked to self-filter for their level of expertise in ML, using the guideline that they should be \textit{``familiar with implementing machine learning models but not an expert, such as if you have taken an introductory course to machine learning or you are self-taught and have implemented ML projects"}. 
Participants are told that they would be debugging a faulty machine learning script with the help of ChatGPT, while\textit{ thinking aloud} about their process.
The study time was 60 minutes and participants were compensated \$20 CAD (approximately ~\$14.80 USD at the time of the study) for their participation.  All study procedures are approved by the Research Ethics Board at the institution. 

\subsection{ML Debugging Task}

While there are many complex LLM-assisted tasks that we can consider, ML debugging is unique for several reasons. With the growing popularity of machine learning and AI in the mainstream, more and more hobbyists and workers without formal education or training are learning ML implementation \cite{cai2019software}. As such, the ubiquity of this growing group of ML novices makes this domain novel and impactful to examine. 

Furthermore, ML debugging is also unique from code debugging, where the latter is a more structured process that tends to have a one-to-one mapping of problems and solutions. Code fixes are also generally directly verifiable through compilation and passing unit tests. Conversely, issues in ML code can compound together and interact \cite{nahar2023meta}. The solution is also harder to verify, as the ideal performance of a model on a dataset is usually not known. To successfully solve them, it can involve incorporating both conceptual knowledge and looking at empirical impacts on the performance to make adjustments to the ML code. This requires a robust mental model of ML debugging, including a) diagnosing the cause(s), b) identifying potential solutions, and c) validating the improvement in performance. For a novice without sufficient expertise, this can be very challenging and warrants assistance. 

For the experiment, we develop a simple ML training script using the UCI \textit{Adult Income} dataset \cite{adult_2}, 
with the task of predicting if a person is high income ($y=1$) or not ($y=0$). The code, written with Python libraries in a Google Colab, includes processing the data, creating the train/test datasets, fitting a \texttt{scikit-learn} \textit{Random Forest} (RF) classifier, and evaluating the performance (F1 score and other metrics). We artificially introduce three problems into the code that would cause noticeable performance degradations on the test performance, see Table \ref{tab:errors} for their details and solutions. \textbf{E1} and \textbf{E2} reflect `code bugs', while \textbf{E3} is a result of the distribution of the dataset. The participant is not required to solve the errors using the provided solutions, as we accept any attempts at applying \textit{conceptually sound} methods to \textit{correctly-identified} problems.

\begin{table}[t!]
\caption{Summary of the three problems introduced in the machine learning debugging task and their solutions.}
\label{tab:errors}
\vspace{-1em}
\begin{tabular}{m{0.04\linewidth}m{0.48\linewidth}m{0.35\linewidth}}
\toprule
\textbf{ID} & \textbf{Error Description} & \textbf{Example Solution} \\
\midrule
\textbf{E1} & Overfitting of the Random Forest model due to complexity hyperparameters not being set. & Limit model complexity hyperparameters, such as \texttt{max\_depth=5}. \\
\midrule
\textbf{E2} & Data distribution shift due to the training/test datasets being unshuffled when split. & Set \texttt{shuffle=True} in \texttt{train\_test\_split().} \\
\midrule
\textbf{E3} & Data imbalance in the dataset which causes poor performance in the minority class. & Set model parameter \texttt{class\_weight = balanced}, or resample data with SMOTE.\\
\bottomrule
\end{tabular}
\end{table}

\subsection{Study Procedure}

In the \textbf{Pre-Task}, participants filled out a survey to self-rate their experience with machine learning and using LLMs, as well as five \textit{knowledge-based} multiple-choice questions on basic machine learning concepts --- this is to estimate their actual abilities for the task, as self-reports may be biased. 

At the start of the \textbf{Main Task}, participants were shown a tutorial demonstrating the \textit{think-aloud} protocol if they weren't already familiar with it. They then received an explanation of the machine learning code and the objective of finding and fixing errors, although they were not told the total number of errors or what metrics they should focus on. They were given free reign to use ChatGPT 4o-mini\footnote{We use the \textit{mini} version as it was available on free accounts and we did not necessarily require a state-of-art model for the study.} to complete the task. The participants' workflows and queries were closely monitored and recorded by screen and audio. After the session, their modified code is scored on held out dataset and the number of problems they fixed were counted. Their transcripts with ChatGPT were also saved for further analysis. 

In the \textbf{Post-Task} survey, participants rated their experience using ChatGPT for the task and any perceived improvements in their ML skills on a 7-point Likert scale. Some of the subjective perception questions are adapted from the UMUX questionnaire \cite{finstad2010usability}. Lastly, participants answered semi-structured interview questions, focusing on topics such as their strategy for using ChatGPT when debugging, the quality of their workflow, and what an ideal ML debugging tool might look like.

\section{Results}
\label{sec:results}

We describe our findings in terms of task performance, workflows (reliance actions and outcomes), and subjective perceptions.

\subsection{Task Performance}
Our analysis first examines the participants' ChatGPT-assisted performance in correcting the errors. Everyone identified and solved (or attempted to solve) at least one problem, with four participants solving two, and two participants solving all three.  
The best possible F1 score on the holdout dataset using an RF model is around 0.32, as experimentally determined by the research team. The participants achieved scores ranging 0-0.28, with 0.16 being the performance of the unmodified code. Due to the sensitivity of ML code, it was possible to implement a partial solution but still severely degenerate the performance.

Interestingly, we find that participants' holdout F1 scores are correlated to their performance on the initial ML knowledge quiz (but not their self-reported ML experience), with Pearson's $r=.93$, $p<.001$. As we have a small number of participants, we do not assign any strong implications to these results. However, taken together with prior research, this suggests that less skilled users struggle with using LLMs to \textit{accelerate} their implementation. For a task as complex as machine learning, strong mental models of the domain is necessary to appropriately oversee ChatGPT's outputs.

\subsection{Reliance Actions and Outcomes}

While ChatGPT helped everyone to some extent, we observed that participants' reliance \textit{actions} and \textit{outcomes} had some important differences. We define the relevant terms as follows: 
\begin{enumerate}
    \item \textbf{Reliance Action:} This describes how the participant worked with the LLM -- either they were \textit{leading the LLM} through asking specific, and planned-out questions; or they were \textit{led-by the LLM} by asking open-ended questions and following the LLM's suggestions. While subtle, this separates \textit{who} (the user, or the LLM) is the primary driver of debugging.   
    This dichotomy of behaviour is related to the \textit{acceleration} vs \textit{exploration} paradigm defined by \citeauthor{barke2023grounded} and the \textit{shepherding} vs \textit{drifting} behaviour documented by \citeauthor{prather2023s}. 
    \item \textbf{Reliance Outcome:} Using the Appropriate Reliance framework \cite{schemmer2023appropriate, bo2024rely}, we define four categories of outcomes in terms of if the user relied on the LLM and if they made the correct decision -- \textit{Rely on self} (when the LLM is wrong), \textit{Rely on LLM} (when the LLM is correct), \textit{Over-rely} (relying on LLM when it is wrong), and \textit{Under-rely} (not relying on LLM when it is right). The first two demonstrate correct reliance, while the latter two are incorrect reliance.
\end{enumerate}

\begin{figure}[t!]
    \centering
    \includegraphics[width=\linewidth]{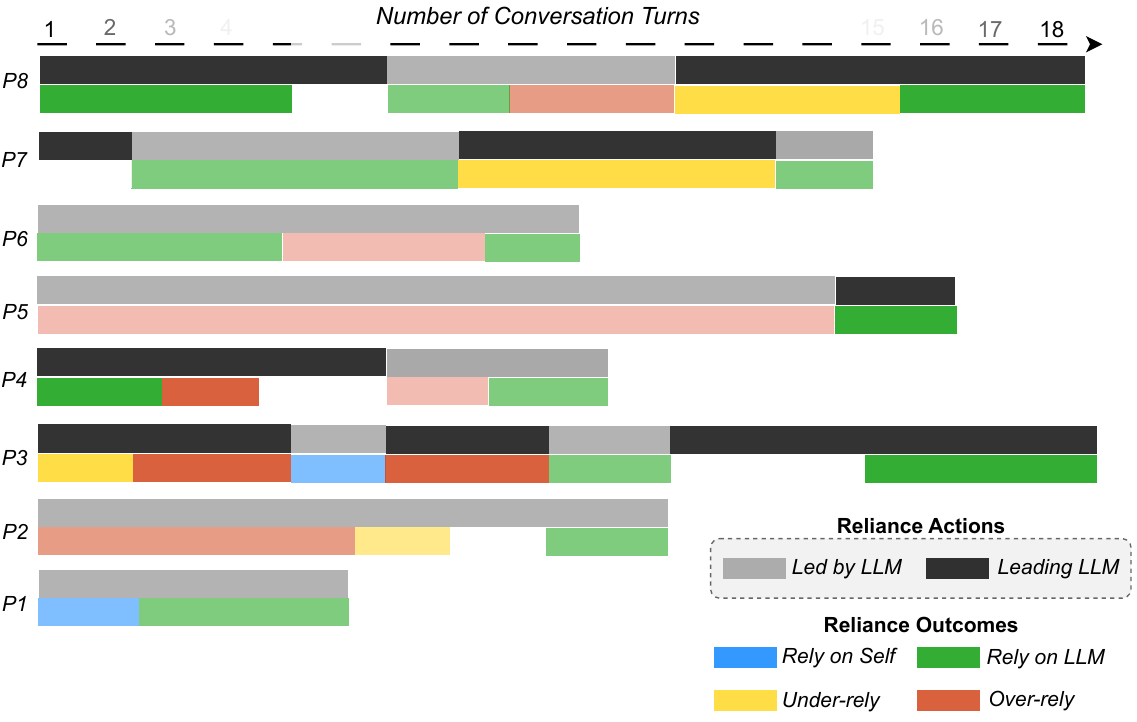}
    \vspace{-1em}
    \caption{Workflow analysis for each participant with reliance outcomes (top bar) and corresponding LLM actions (bottom bar). To disambiguate between the \textit{leading/led-by} paradigm, the reliance outcome colours when the user \textit{lead} the LLM are more saturated than when \textit{led-by} the LLM.}
    \vspace{-1em}
    \label{fig:workflow}
\end{figure}

\textbf{Reliance Actions} were coded for each conversation turn between the participant and ChatGPT, informed by notes and audio transcripts taken during the study session. For example, a \textit{leading} participant independently hypothesized the causes of an error and asked, \textit{``What are the most important hyper parameters for Random Forest?"}; while a \textit{led-by} participant queried, \textit{``What is wrong with this code?"} and verbalized they were unsure of how to start. 
\textbf{Reliance Outcomes} were coded based on what the participant did with ChatGPT's response --- either applying the suggestions or moving on. A simplified diagram of the participants workflows is shown as Figure \ref{fig:workflow}. For each participant, the top bar indicates their reliance actions, and the bottom bar indicates the  outcomes.

Participants who asked for general guidance (\textit{led-by}) and shifted the workload to ChatGPT generally performed worse than participants who used ChatGPT deliberately to support their debugging plans (\textit{leading}). 
Four of the five top scorers on the holdout dataset were \textit{leading} ChatGPT in half or more of the conversation turns (\textbf{P3}, \textbf{P4}, \textbf{P7}, and \textbf{P8}). Qualitatively, we observe that \textit{under-reliance} occurred more when the participants were controlling the debugging; while \textit{over-reliance} is more frequent when the participant blindly followed the LLM. This is likely explained by less experienced novices lacking relevant knowledge to drive decisions in the task. We also note that amongst the \textit{led-by} participants, some seemed primarily driven by their high trust in ChatGPT -- such as \textbf{P5} and \textbf{P6} who presented as power LLM users. 

There are novice-specific factors that degraded interactions with ChatGPT.
We label several types of meta-cognitive errors related to how a lack of robust mental models interacts with the compliant nature of LLMs: 
\begin{itemize}
    \item \textbf{Leading Query}: Participants who know what they want to explore often inject `leads' into their queries, which was problematic if the idea was wrong --- such as, \textit{``Give me the code for feature standardization"}, where standardization is not necessary for RF models. ChatGPT would provide the code and lead the participant into an unnecessary over-reliance.
    \item \textbf{Filtering}: When asked broad questions, ChatGPT would generate a long list of potential solutions (sometimes more than 10 items). This was very difficult for users to filter through in the time constraint of the task, and sometimes lead to under-reliance. 
    \item \textbf{Verification}: Even if the participant identifies the correct actions to take and ChatGPT provides reasonable advice, small differences in the suggestions can drastically impact the outcome. \textbf{P8} spent significant time searching for the best RF hyperparameter values without success because ChatGPT's suggested values were not optimal for the dataset. 
    More generally, many participants exhibited difficulty with understanding which ML metrics to pay attention to (training vs testing performance; accuracy vs F1 score) and had a hard time with communicating changes in performance to ChatGPT.
\end{itemize}

\subsection{Subjective Perceptions}
Despite the difficulty of the task, overall perceptions of ChatGPT’s ability to aid in the task are positive based on the post-task survey. All participants indicated they believed they learned something through using ChatGPT, although their confidence in their ML debugging skills did not improve. Based on interview results, learning mostly referred to low-level and syntactic information -- such as the hyperparameters in an RF model -- \textit{not} mental models of ML debugging. 
Perceived usability and success with prompting is more mixed, and several participants indicated in the post interview that they would take different prompting strategies (like adding context and specificity) if they were to redo the task (\textbf{P1}, \textbf{P2}, \textbf{P4}, \textbf{P5}, \textbf{P8}). 

In terms of the generated content, participants commented that ChatGPT's responses were too broad and lacked context (\textbf{P1}, \textbf{P2}, \textbf{P7}) and the amount of information is overwhelming to filter (\textbf{P3}, \textbf{P8}) -- signaling cognitive overload. However, many also liked that ChatGPT is more selective than Google Search, which helped them find relevant information more efficiently (\textbf{P3}, \textbf{P6}, \textbf{P7}). Some indicated they would like to see suggestions of common errors to debug as a general guideline (\textbf{P3}, \textbf{P4}). 
\textbf{P6}, who was the only one who faced issues with running the ChatGPT generated code, wanted verification that the syntax can compile. These perception demonstrate the potential for ChatGPT to guide users in complex tasks, but also the some shortcomings of vanilla ChatGPT’s interaction strategies for task novices. 

\section{Discussion}
\label{sec:discussion}

\subsection{Key Findings}
We present the preliminary findings of novice-LLM workflows in a highly challenging task with high requirements for domain knowledge -- machine learning debugging. Participants who were successful in the task tended to have better mental models and were able to \textit{lead} the debugging process, while less experienced participants were more likely to delegate the bulk of the work and be \textit{led-by} ChatGPT. We also identify some interesting modes of interaction failure.
When asked open-ended queries, ChatGPT would present a breadth of ideas but not pinpoint the most salient suggestions, leading to cognitive overload and under-reliance. On the other hand, users with incorrect hypotheses about the task asked poorly phrased queries to ChatGPT, leading to `rabbitholes' of over-reliance on irrelevant advice. 
%

\subsection{Implications on Cognition}
Designing LLMs to collaborate with novices is challenging \edit{in many ways. The interactions between the user's meta-cognition, mental models, and the LLM's behaviour can result in various pitfalls, some of which were documented in the study.} While an expert may be able to leverage the LLM using precise prompting and seasoned verification skills to automate the task, we observed that novices struggled with both aspects of task performance and learning. \edit{There may even be further tensions between optimizing the debugging task performance and having the novice user meaningfully engage and \textit{lead} the task --- as in, even when novices \textit{are} engaged, their faulty mental models of the task can be reinforced through the LLM's sycophancy, resulting in poor learning outcomes. Beyond in-task over-reliance errors, the reduction of high-quality learning in the long term, especially without proper educational scaffolding, is a key concern in novice-LLM interactions. }

 An important question posed for future research is how the LLM interaction paradigm can be augmented to correctly engage novice users who may have insufficient experience or even incorrect beliefs about the task they are working on. Could LLMs that generate satisficing solutions lead to a culture of novice implementers with poor understanding of how and why the solution was constructed? Will there be downstream effects of improper reliance on LLMs in the educational context, where students bypass foundational learning or are reinforced with bad mental models? We propose some ways to improve novice-LLM interactions in the next section.


\subsection{Improving Novice-LLM Interactions}

We suggest some ways for how novice-LLM interactions can  be augmented to avoid the pitfalls observed in our study. We divide these ideas based on \textit{which side} of the conversation they take effect:
\begin{enumerate}
    \item \textbf{LLM-side improvement}: ChatGPT's propensity to answer questions complacently and agreeably, even if the user's questions are uninformed, made it a bad guide for novices. \textit{Sycophancy}, a trait that is learned through training by reinforcement learning on human feedback, describes how LLMs would align themselves to reflect the users beliefs, even if the beliefs are incorrect \cite{sharma2023towards}. While we did not observe outlandishly sycophantic statements, ChatGPT was not firm with correcting users' faulty mental models and was overall ineffective at helping novices improve their task knowledge. Adapting LLMs to be more proactive in correcting user's faulty assumptions can help steer novices onto the right track \edit{and engage in appropriate reflection to form correct hypotheses in the future}.
    
    \item \textbf{User-side improvement}: \edit{Users also face challenges as a result of struggling to articulate their intent in their prompts}  \cite{zamfirescu2023johnny, tankelevitch2024metacognitive}. Some of the participants in the study also expressed wanting to see examples of good queries. Given that we conducted the experiment with mostly Computer Science students, who should have higher technical expertise and AI literacy than the general public, we expect that lay users would need even more support. Providing a tutorial of appropriate usage and effective prompting strategies may help user ask higher quality questions \edit{and guide them away from being led into over-reliance} \cite{ma2024you}.
    
    \item \textbf{Bidirectional improvement}: Given that both the user and LLM participate in the conversation, improvements can be made from both sides, iteratively.
    Since the mental model of a novice can be highly flawed due to lack of information, incorrect information, or incorrect understanding of processes, it would benefit the LLM to customize its response to tackle these misconceptions \cite{chandra2024watchat}. \textit{Mutual theory of mind} is a cognitive psychology construct that has been explored in the LLM space, describes the ability of humans to infer what information another person knows. It has been applied as a framework to improving trust and perceptions in human-LLM collaboration tasks \cite{wang2021towards, zhang2024mutual}. 
\end{enumerate}

We propose these future directions for the research community as ways to improve interactions between novices and their LLM assistants, \edit{with a focus on reducing over-reliance, enhancing cognitive engagement, and improving mental models of the task.} 

\clearpage
\bibliographystyle{ACM-Reference-Format}
\bibliography{references}

\end{document}